\documentclass[twocolumn,english,aps,prd,reprint,groupedaddress,notitlepage,nobibnote,nofootinbib,preprintnumbers,showpacs]{revtex4-1}
\pdfoutput=1
\usepackage{lmodern}

\usepackage[T1]{fontenc}
\usepackage[latin9]{inputenc}
\usepackage{geometry}
\usepackage{color}
\usepackage{babel}
\usepackage{amsmath}
\usepackage{amssymb}
\usepackage{graphicx}
\usepackage{esint}
\usepackage{wasysym}
\usepackage{array}
\usepackage{comment}
\usepackage{tabu}
\usepackage{relsize}
\usepackage{pifont}

\usepackage{bm}
\usepackage{subfigure}
\usepackage{slashed}
\usepackage{epsfig}
\usepackage{hyperref}
\usepackage{color}
\usepackage{amsmath}
\usepackage{bbold}

\usepackage[table]{xcolor}

\hypersetup{
 citecolor=blue,linkcolor=blue,urlcolor=blue}
\makeatletter

 \@ifundefined{textcolor}{}
 {
   \definecolor{BLACK}{gray}{0}
   \definecolor{WHITE}{gray}{1}
   \definecolor{RED}{rgb}{1,0,0}
   \definecolor{GREEN}{rgb}{0,1,0}
   \definecolor{BLUE}{rgb}{0,0,1}
   \definecolor{CYAN}{cmyk}{1,0,0,0}
   \definecolor{MAGENTA}{cmyk}{0,1,0,0}
   \definecolor{YELLOW}{cmyk}{0,0,1,0}
 }
\usepackage{babel}

\newcommand{\kev}{\text{keV}}

\newcommand{\gev}{\text{GeV}}
\newcommand{\tev}{\text{TeV}}

\renewcommand{\eqref}[1]{Eq.~(\ref{#1})}

\newcommand{\secref}[1]{Sec.~\ref{sec:#1}}

\newcommand{\figref}[1]{Fig.~\ref{fig:#1}}

\newcommand{\tableref}[1]{Tab.~\ref{table:#1}}
\newcommand{\tablesref}[2]{Tabs.~\ref{table:#1} and \ref{table:#2}}

\newcommand{\gsim}{\lower.7ex\hbox{$\;\stackrel{\textstyle>}{\sim}\;$}}
\newcommand{\lsim}{\lower.7ex\hbox{$\;\stackrel{\textstyle<}{\sim}\;$}}

\newcommand{\bea}{\begin{eqnarray}}
\newcommand{\eea}{\end{eqnarray}}

\newcommand{\Eq}[1]{Eq.~(\ref{#1})}

\newcommand{\Sec}[1]{Sec.~\ref{#1}}

\newcommand{\cmark}{\text{\ding{51}}}%
\newcommand{\xmark}{\text{\ding{55}}}%

\newcommand{\sslash}[1]{\ensuremath\raisebox{-0.00cm}{{\small\slash}}\hspace{-0.21cm}#1\/}

\newcommand{\chiDM}{X}
\newcommand{\Uc}{U}
\newcommand{\Dc}{D}
\newcommand{\Ec}{E}

\newcommand{\Dcbar}{\bar D}

\newcommand{\mdm}{M}
\newcommand{\yes}{\textcolor{green}{$\cmark$}}
\newcommand{\yess}{\textcolor{green}{$\cmark$}$^*$}
\newcommand{\no}{\textcolor{red}{$\xmark$}}

\begin{document}

\preprint{\hbox{CALT-2014-034} }

\title{Effectively Stable Dark Matter}

\author{Clifford Cheung}
\author{David Sanford}
\affiliation{Walter Burke Institute for Theoretical Physics \\California Institute of Technology, Pasadena, CA 91125}
\date{\today}
\email{clifford.cheung@caltech.edu,dsanford@caltech.edu}
\begin{abstract}

We study dark matter (DM) which is cosmologically long-lived because
of standard model (SM) symmetries.  In these models an approximate stabilizing
symmetry emerges accidentally, in analogy with baryon and lepton
number in the renormalizable SM.  Adopting an effective theory
approach, we classify DM models according to representations of
$SU(3)_C\times SU(2)_L\times U(1)_Y \times U(1)_B\times U(1)_L$,
allowing for all operators permitted by symmetry, with weak scale DM
and a cutoff at or below the Planck scale.  We identify
representations containing a neutral long-lived state, thus
excluding dimension four and five operators that mediate dangerously
prompt DM decay into SM particles.  The DM relic abundance is obtained
via thermal freeze-out or, since effectively stable DM often carries
baryon or lepton number, asymmetry sharing through the very operators
that induce eventual DM decay.  We also incorporate
baryon and lepton number violation with a spurion that parameterizes hard breaking by
arbitrary units.  However, since proton stability precludes certain
spurions, a residual symmetry persists, 
maintaining the cosmological stability of certain DM representations.
Finally, we survey the phenomenology of effectively stable DM as
manifested in probes of direct detection, indirect detection, and proton
decay.

\end{abstract}

\maketitle

\section{Introduction}

Dark matter (DM) is elegantly accounted for by a neutral,
cosmologically long-lived particle beyond the standard model (SM).  In
most circumstances, however, DM stability is ensured by a symmetry
that is simply imposed by fiat.  While as much can be expected from DM
effective theories, similarly ad hoc choices are often needed for their
ultraviolet completions, for instance in theories of supersymmetry or
extra dimensions where $R$-parity or $KK$-parity are assumed.

In contrast, the SM implements stability with less contrivance: charge
stabilizes the electron, while angular momentum stabilizes the
neutrino.  The proton is not guaranteed to be stable, but like DM it
is cosmologically long-lived, with current limits bounding its
lifetime to be greater than $\sim 10^{33}$ years.  Famously, the SM
gauge symmetry explicitly forbids baryon and lepton number violation
at the renormalizable level, suggesting an argument for proton
stability from effective theory.  This mechanism, whereby an
approximate symmetry arises as the byproduct of existing symmetries,
is sometimes referred to as an accidental symmetry.

In this paper we argue that DM, like the proton, can be cosmologically stable as an accident of SM symmetries.  For our analysis we consider the symmetry group
\begin{eqnarray}
SU(3)_C \times SU(2)_L \times U(1)_Y
\times U(1)_B\times U(1)_L,
\end{eqnarray}
which is exactly preserved in the renormalizable SM to all orders in
perturbation theory.  Beyond the renormalizable level, $B$ and $L$ may
be approximate or exact, depending on whether they are gauged in the
ultraviolet.  While  SM quark and lepton flavor
are approximate symmetries, we will not consider them here.

In our analysis, we enumerate models according to the quantum numbers
of DM under the SM symmetry group.  We take the stance of effective
theory throughout: all interactions, renormalizable and
non-renormalizable, are to be included with order one coefficients.  We
assume a DM mass $\mdm$ near the weak scale and an effective theory
cutoff $\Lambda$ at or below the Planck scale.

To start, we discard all representations without a component neutral under color and electromagnetism.  We  then discard all 
representations that permit DM decay into SM particles via dimension
four or five operators.  In such cases even a Planck scale cutoff is
insufficient to prevent cosmologically prompt DM decay.  On the
other hand, decays induced by dimension six operators hover right at the
boundary of current bounds from indirect detection, assuming a cutoff
near the scale of grand unification (GUT) \cite{Arvanitaki:2008hq}.  

Applying these criteria, we enumerate models of effectively
stable DM, focusing on all possible fermionic
and scalar DM candidates whose leading decays occur at dimension six
or seven.  These results are summarized in
\tablesref{constraintstable_fermiondim6}{constraintstable_scalardim6}
for dimension six and
\tablesref{constraintstable_fermiondim7}{constraintstable_scalardim7}
for dimension seven decays, respectively.

Our approach resonates with that of minimal
DM~\cite{Cirelli:2005uq,Cirelli:2007xd} but with the crucial
difference that we incorporate both $B$ and $L$ and faithfully assume the presence of all
interactions not forbidden by symmetry.  Without $B$ and $L$, representations smaller than the quintuplet of $SU(2)_L$, DM will
promptly decay to SM particles via renormalizable interactions \cite{Cirelli:2005uq,Cirelli:2007xd,DiLuzio:2015oha}.  This
is consistent with our own finding that effectively stable DM in small
representations of $SU(2)_L$ must carry $B$ or $L$.  For exactly this
reason many of these models may be generated through the mechanism of asymmetric DM \cite{Kaplan:2009ag, Zhao:2014nsa}.  Previous authors have also considered accidental stabilization of DM by flavor symmetries \cite{Lavoura:2012cv} or new gauge symmetries \cite{Antipin:2015xia}.

Finally, we consider the possibility that $B$ and $L$ are merely approximate.  For this analysis we introduce a spurion parameterizing the 
hard breaking of baryon and lepton number by arbitrary units $\Delta B$ and $ \Delta L$, respectively.  In many models
of effectively stable DM, this induces prompt decays.  However, not
all values of $(\Delta B,\Delta L)$ are phenomenologically safe: the
non-observation of proton decay suggests that dimension six operators
of the form $Q^3L$ should be forbidden, thus precluding hard breaking
by units of $(\Delta B,\Delta L)=(1,1)$.  Remarkably, given arbitrary breaking by any $(\Delta B,\Delta L)\neq(1,1)$, we still maintain a handful of viable candidates for effectively stable DM.  In
these models, DM is long-lived because of the SM gauge
symmetry together with the stability of the proton.

The remainder of this paper is as follows.  In \secref{model} we
enumerate viable models of effectively stable DM using the criteria described above.  In \secref{phenom} we study
experimental constraints on these theories from direct detection,
indirect detection, and proton decay.  Finally,  we
summarize our results and discuss future directions in \secref{conclusion}.

\section{Classification of Models}

\label{sec:model}

In this section we enumerate representations of the SM symmetry group with a DM component that is neutral, cosmologically long-lived, and
generated with the observed relic abundance.  We adopt an effective
theory perspective in which any operator allowed by symmetry is
present with a strength set by a sub-Planckian cutoff.

\subsection{Neutrality}

The quantum numbers of DM are parameterized by a discrete choice of
representations for $SU(3)_C$ and $SU(2)_L$ together with a continuous
choice of charges for $U(1)_Y$, $U(1)_B$, and $U(1)_L$.  For
simplicity, we focus on pure gauge eigenstates, assuming that DM is
either a complex scalar or a Dirac fermion.

To begin, we restrict to representations that have a neutral component
under the unbroken SM gauge symmetry.  Thus, DM is an $SU(3)_C$
singlet.  For DM that is a $k$-plet under $SU(2)_L$, we require that
$k \ge 2 |Y| + 1$, where the hypercharge $Y$ is quantized to an
integer or half-integer value if $k$ is odd or even, respectively.

Charge neutrality places no constraints on the $B$ or $L$ charges of
DM.  While irrational values of $B$ and $L$ are allowed a priori,
this is literally equivalent to enforcing DM number as an exact
symmetry of the Lagrangian.  Furthermore, exact global
symmetries are known to conflict with black hole no-hair
theorems~\cite{Banks:2010zn}.  For these reasons we restrict to rational
values $B$ and $L$.

While $B$ and $L$ are symmetries of the renormalizable SM, they may of course
be spontaneously or explicitly broken in the full theory.  To 
parameterize these effects conservatively, we introduce effective hard breaking of
$B$ and $L$ into the low energy theory with a dimensionless spurion for symmetry
breaking by units of $(\Delta B, \Delta L)$.  For remainder of this section we assume that $(\Delta B,\Delta L)=(0,0)$, but return to the issue of explicit breaking later on in \Sec{sec:phenom}.

\subsection{Stability}

We now determine the leading operators that are allowed by symmetry and
mediate DM decay.  An operator that induces DM decay takes the form
\begin{eqnarray}
\mathcal{O}_{\rm DM} & = &  
\chiDM \mathcal{O}_{\mathrm{SM}} \, ,
\end{eqnarray}
where $X$ is the fermion or scalar field that contains DM and
$\mathcal{O}_{\mathrm{SM}}$ is an operator composed entirely of SM fields.
For later convenience, we define
\begin{eqnarray}
N &=& [{\cal O}_{\rm DM}] \, ,
\end{eqnarray}
to be the dimension of the DM decay
operator.
The quantum numbers of $\chiDM$ are equal and opposite to
those of $\mathcal{O}_{\mathrm{SM}}$, so to enumerate all decay
operators it suffices to determine all operators $
\mathcal{O}_{\mathrm{SM}}$ of a given charge and operator dimension.

We define a fiducial decay rate into SM particles,
\begin{eqnarray}
\Gamma \left( \chiDM \rightarrow \mathrm{SM} \right) & \sim & \frac{\mdm}{4\pi}
 \left(\frac{\mdm}{\Lambda} \right)^{2(N-4)}\,,
\label{eq:Gammafid}
\end{eqnarray}
corresponding to two-body decay via ${\cal O}_{\rm DM}$. Depending
on the precise operator, this decay may be three-body or higher.
Moreover, decays into SM fermions will involve flavor structures that may further suppress the width.  In any case, the fiducial decay rate
in \Eq{eq:Gammafid} should be taken as an overestimate.

By definition, a cosmologically stable DM particle $\chiDM$ has a
lifetime of order the age of the universe,
\begin{eqnarray}
\tau(\chiDM \rightarrow \textrm{SM}) & \gtrsim & 10^{18}
\textrm{ sec}\quad (\textrm{age of universe})\,.
\end{eqnarray}
However, this bound is weaker than experimental constraints on
cosmic ray production from DM decay into
positrons~\cite{Ibarra:2013zia}, gamma rays~\cite{Ando:2015qda},
antiprotons~\cite{Giesen:2015ufa}, and neutrinos~\cite{Rott:2014kfa}.
These limits all place similar constraints on the DM lifetime, of
order
\begin{eqnarray}
\tau(\chiDM \rightarrow \textrm{SM}) & \gtrsim & 10^{26} \textrm{ sec}\quad (\textrm{indirect detection})\,,
  \label{eq:ID}
\end{eqnarray}
for $\mdm$ of order the weak scale.  While CMB bounds are also
stronger the one from the age of the universe, they are still weaker
than indirect search bounds by $\sim 2-3$ orders of
magnitude~\cite{Cline:2013fm}.

Comparing  \Eq{eq:ID} 
to \Eq{eq:Gammafid}, we obtain
upper bounds on the cutoff of higher dimension operators.  For
dimension five, six, and seven operators, this implies a schematic
lower bound on the cutoff,
\begin{eqnarray}
\Lambda & \gtrsim & \left\{
\begin{array}{cc}
\left( \cfrac{\mdm}{1~\tev}
\right)^{\frac{3}{2}} 10^{29}~\gev\,, & N=5 \\
 \left( \cfrac{\mdm}{1~\tev}
\right)^{\frac{5}{4}} 10^{16}~\gev\,, & N=6\\
 \left( \cfrac{\mdm}{1~\tev}
\right)^{\frac{7}{6}} 10^{12}~\gev\,, & N=7
\end{array} \right. .
\end{eqnarray}
Since \Eq{eq:Gammafid} is an overestimate, this bound on the cutoff is
conservative.

\begin{figure}[t]
\hspace*{-0.25in}
    \includegraphics*[width=0.48\textwidth]{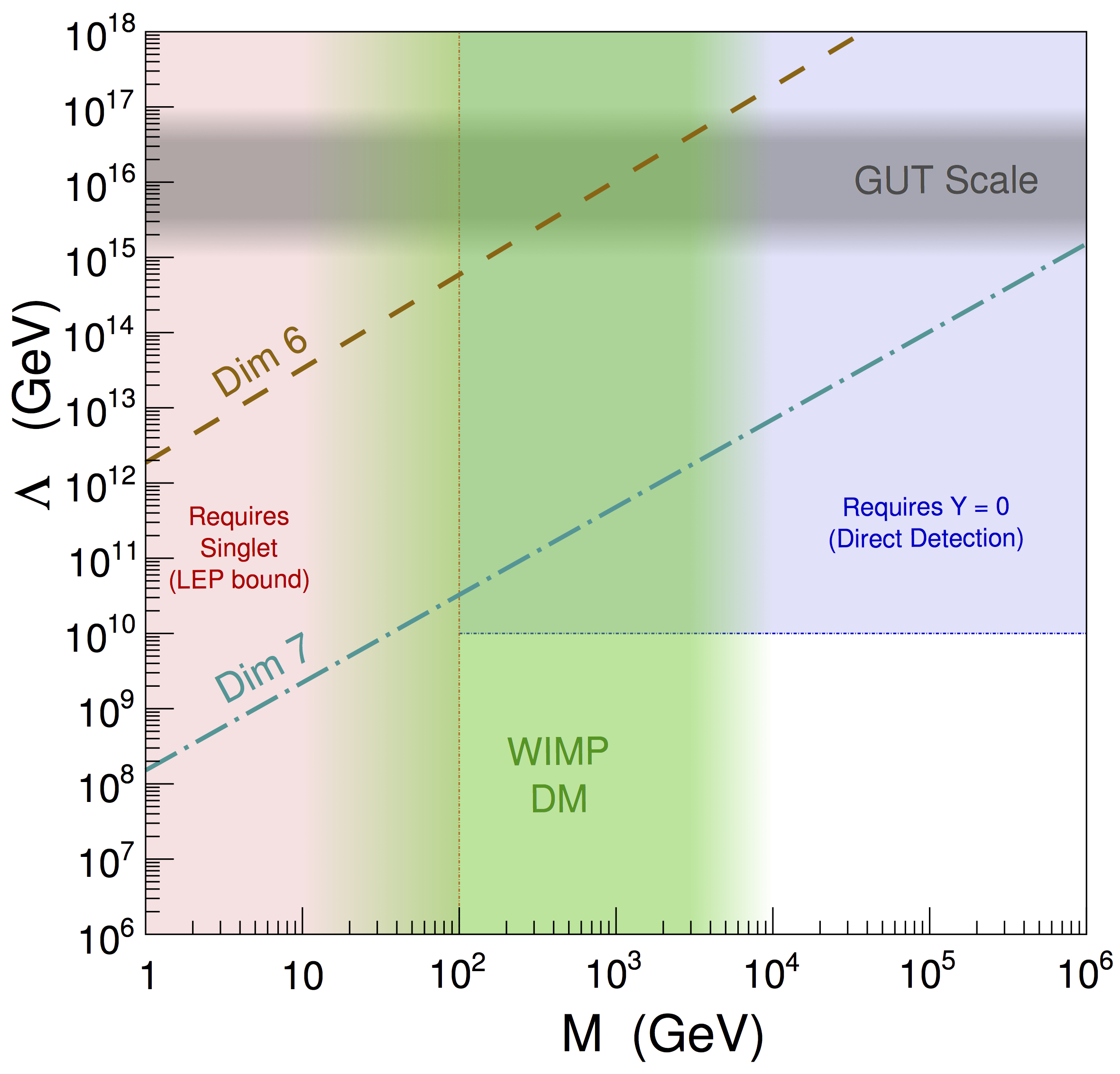}
\caption{\label{fig:LambdaLimits} Lower bound on the effective theory
  cutoff $\Lambda$ as a function of the DM mass $M$ in order to avoid
  cosmologically prompt decay via dimension six (gold dashed) and
  seven (teal dot-dashed) operators.  LEP (red shaded) excludes light, weakly
  interacting or hypercharged DM, while direct detection (blue
  shaded) excludes hypercharged DM that cannot accommodate an
  inelastic splitting to evade bounds.}
\end{figure}

For dimension five operators, a sub-Planckian cutoff $\Lambda$ would require $\mdm \lesssim 10~\kev$ for DM, so hereafter we
consider only DM decay via dimension six and seven operators.
Notably, dimension six operators are a particularly intriguing portal
for weak scale DM because GUT-suppressed dimension six operators
induce decays that lie just at the boundary of current experimental
limits.  For dimension seven operators, the bound on $\Lambda$ is even
smaller.  \figref{LambdaLimits} shows the lower bound on $\Lambda$ as
a function of $M$ for dimension six and seven decay operators.  The
bounds for dimension five operators are not shown because they require a
cutoff above the Planck scale.  Also shown are the natural scale for
DM mass $M$ near the weak scale and cutoff scale $\Lambda$ at the GUT
scale, as well as regions excluded by experimental limits from LEP and
direct detection, to be discussed later.

To summarize, a sub-Planckian cutoff implies that cosmologically stable,
weak scale DM forbids all dimension four or five decay
operators.  This is a stringent constraint on the DM representation.
For example, since a pure singlet scalar $X$ can couple to any
dimension four operator ${\cal O}_{\rm SM}$ in the SM, the associated
DM will promptly decay at dimension five.  At the opposite extreme is
scalar or fermionic DM with extremely large charges.  In this case
lower dimensional operators are forbidden simply because so many SM
fields have to be included in the decay operator just to preserve the
symmetry.  As noted in \cite{Cirelli:2005uq}, this occurs for
$k$-plets of $SU(2)_L$ with large values of $k$.

All of our fermionic DM candidates and most of our scalar DM
candidates carry non-zero $B$ or $L$.  The reason is that most
representations with zero $B$ or $L$ typically have very low dimension
decay operators.  For example whenever ${\cal O}_{\rm SM}$ is a SM fermion bilinear that is
$B$ and $L$ neutral,  it is always possible to construct a lower dimension operator via
replacements of the form $Q \Uc \rightarrow H^\dagger$, $L \Ec
\rightarrow H$, $\bar{Q} \bar\sigma^\mu Q \rightarrow B^\mu, W^\mu$,
etc.  As a result, viable DM tends to carry $B$ or $L$, or reside in a large representation of $SU(2)_L$ so that gauge invariance requires many Higgs fields in the operator.

Large hypercharge can serve the same role in terms of stability, but
as noted earlier, a neutral DM component requires that $k \ge 2 |Y| +
1$ for a $k$-plet under $SU(2)_L$.  This in turn bounds the net
$SU(2)_L \times U(1)_Y$ charges of the SM fields that couple to
$\chiDM$.  In particular, it largely excludes decay operators involving
$\Ec$, since this SM field has a sizable hypercharge and no $SU(2)_L$
charge.

\tablesref{constraintstable_fermiondim6}{constraintstable_scalardim6}
list all fermionic and scalar DM representations, respectively, whose leading decay is mediated by a
dimension six operator.  Every representation is color neutral and
every representation carries $B$ or $L$ except the scalar $SU(2)_L$
sextet.  For the fermionic DM we forbid both members of the Dirac pair
from decaying via dimension five or lower operators.  Also, shown are
various attributes of the model regarding direct detection and
explicit $B$ and $L$ breaking, to be discussed later.
\tablesref{constraintstable_fermiondim7}{constraintstable_scalardim7}
are the same as
\tablesref{constraintstable_fermiondim6}{constraintstable_scalardim6}
except they apply to DM representations whose leading decay is
mediated by a dimension seven operator.

\begin{table}[t]
\begin{center}
Fermionic DM $(N=6)$
\end{center}
\begin{tabular}{|c|c|c|c|c|c|c|}
\hline
 $SU(2)_L$ & $U(1)_Y$ & $U(1)_B$ & $U(1)_L$ & $\sslash{B},\sslash{L} $& $\sigma_{\rm SI}$  & ${\cal O}_{\rm DM}$ \\
 \hline
 1 & 0 & 1 & 0 & \yes & \yes  & $\Uc \Dc^2 \chiDM$ \\
 2 & $-1/2$ & $-1$ & 0 & \no & \no  & $Q^3\chiDM$ \\
 2 & 1/2 & $-1$ & 0 & \no & \no  & $Q \Dcbar^2 \chiDM$ \\
 3 & $-1$ & 0 & $-1$ & \no & \no  & $H^3 L \chiDM$ \\
 3 & 0 & 1 & 0 &  \no & \yes  & $\bar{Q}^2 \Dc \chiDM$ \\
 3 & 1 & 1 & 0 & \yes & \no  & $\bar{Q}^2 \Uc \chiDM$ \\
 4 & $-1/2$ & $-1$ & 0 & \no & \no  & $Q^3\chiDM$ \\
 4 & 3/2 & 0 & $-3$ & \no & \no  & $L^3\chiDM$ \\
 5 & $-1$ & 0 & $-1$ & \yes & \no  & $H^3 L \chiDM$ \\
 5 & 0 & 0 & $-1$ & \yes & \yes  & $H^\dagger H^2 L \chiDM$ \\
 5 & 1 & 0 & $-1$ & \yes & \no  & $H^{\dagger 2} H L \chiDM$ \\
 5 & 2 & 0 & $-1$ & \yes & \no  & $H^{\dagger 3} L \chiDM$ \\

\hline

\hline
\end{tabular}
\caption{\label{table:constraintstable_fermiondim6} Classification of
  fermionic DM which is stable up to dimension six decays.  Columns 1-4
  list DM charges under the SM symmetry group.  Column 5 has a \yes~for DM that remains stable up to dimension six decays even when $B$ and $L$ are explicitly
  broken while preserving proton stability.   Column 6 has a \yes~for DM that is safe from spin-independent DM-nucleon
  scattering through the $Z$ boson.  Column 7 lists an example
  operator mediating DM decay.  }
\end{table}

\subsection{Relic Abundance}

Let us comment briefly on the origin of the DM relic abundance.  DM
with non-trivial SM gauge charges will be equilibrated with the
thermal plasma in the early universe, in which case DM
annihilations are mediated by gauge interactions whose strength is
fixed by the charges of $\chiDM$.  In general there may be additional
interactions between the DM and SM fields, either through the Higgs
boson or through direct couplings to quarks and leptons.  However,
the DM charges are chosen by construction to eliminate renormalizable
decays, so the latter are highly suppressed. On the other hand, the
former are allowed for scalar DM, and from an effective theory
perspective should be included if they are permitted by the symmetries
of the theory.

If gauge interactions dominate DM annihilations, then the relic
abundance is fixed by thermal freeze-out~\cite{Cirelli:2005uq,
  Cirelli:2007xd}.  Since the strength of the $SU(2)_L\times U(1)_Y$
gauge interactions are known, there is only one free parameter
available to tune the relic abundance: the DM mass.  As discussed
in~\cite{Cirelli:2005uq, Cirelli:2007xd}, to obtain DM of the observed
relic abundance today requires $\mdm$ of order the weak scale or
slightly above, depending on the DM spin and quantum numbers.  Of
course, direct couplings to the Higgs~\cite{Cheung:2013dua} can
strongly affect the relic abundance.  However, annihilations mediated
by the Higgs generally add to the DM annihilation cross-section,
depleting the DM abundance and thus requiring a larger DM mass.

Another approach to DM generation is 
asymmetric DM, which fits quite naturally into the effective theory
picture presented here.  In models of asymmetric DM there is an
approximate DM number $X$ that is conserved along with $B$ or $L$.  In
the early universe, some set of interactions, for example higher
dimension operators or sphaelerons, break some combination of $X$,
$B$, and $L$, thus sharing particle asymmetries among the quarks, leptons, and DM.  As we have shown, the majority of effectively stable
DM candidates carry non-zero $B$ or $L$, which implies that the very
same operators that mediate DM decay can also share the asymmetries in
the early universe.  A similar observation was noted about asymmetric
DM models in \cite{Zhao:2014nsa}.  Alternatively, these kinds of
higher dimension operators can also be applied in the reverse direction
to produce the baryon asymmetry \cite{Cheung:2013hza}.

If the decay operator is in chemical equilibrium in the early
universe, then the DM and $B$ or $L$ asymmetries will be shared
according to their charges~\cite{Kaplan:2009ag}.  As is well-known,
for efficient sharing this typically requires light asymmetric DM of
order the $\sim$ GeV scale rather than the weak scale.  For DM that is
a gauge singlet, such as the first entry in
\tableref{constraintstable_fermiondim6}, this offers a viable model of
asymmetric DM, provided additional annihilation modes to deplete the symmetric abundance.  

On the other hand,
models with DM carrying SM gauge charges are excluded by
LEP~\cite{Heister:2002mn,Abdallah:2003xe} for $\sim$ GeV scale masses.
In this case, asymmetric DM is possible only if asymmetry sharing
through the decay operator is inefficient, so the abundance of DM
is less than the amount prescribed by chemical equilibrium, thus
requiring a larger DM mass.

\begin{table}[t]
\begin{center}
Scalar DM $(N=6)$
\end{center}

\begin{tabular}{|c|c|c|c|c|c|c|}
\hline
$SU(2)_L$ & $U(1)_Y$ & $U(1)_B$ & $U(1)_L$ & $\sslash{B},\sslash{L} $& $\sigma_{\rm SI}$  & ${\cal O}_{\rm DM}$ \\
 
\hline
 1 & 0 & 0 & $-2$ & \no & \yes  & $H^2 L^2  \chiDM$  \\
 3 & 0 & 0 & $-2$ & \no & \yes  & $H^2 L^2 \chiDM$  \\
 5 & 0 & 0 & $-2$ &  \no & \yes  & $H^2 L^2 \chiDM$ \\
 5 & 1 & 0 & $-2$ &  \no & \no  & $H^\dagger H L^2  \chiDM$ \\
 5 & 2 & 0 & $-2$ &  \no & \no  & $H^{\dagger 2}L^2  \chiDM$ \\
 6 & 1/2 & 0 & 0 & \yes & \no  & $H^{\dagger 3} H^2 \chiDM$ \\
 6 & 3/2 & 0 & 0 & \yes & \no  & $H^{\dagger 4} H \chiDM$ \\
 6 & 5/2 & 0 & 0 & \yes & \no  & $H^{\dagger 5} \chiDM$ \\
\hline
\end{tabular}
\caption{\label{table:constraintstable_scalardim6} Classification of
  scalar DM which is stable up to dimension six decays.  Notation from
  \tableref{constraintstable_fermiondim6}.}
\end{table}

\section{Experimental Constraints}

\label{sec:phenom}

In this section we survey the bounds on effectively stable DM from
laboratory and telescope experiments.  As we will see, bounds from
direct detection and proton decay have an interesting connection to
explicit breaking of $B$ and $L$.

\subsection{Direction Detection}

\label{sec:directdetection}

Experimental bounds on spin-independent DM-nucleon scattering are
extremely stringent. In particular, Dirac or complex scalar DM with a
spin-independent coupling to the $Z$ boson is excluded by many orders
of magnitude.  Such an interaction arises when $Y \neq 0$, so direct
detection constraints are trivially evaded by DM with vanishing
hypercharge.

 In
\tablesref{constraintstable_fermiondim6}{constraintstable_scalardim6}
and
\tablesref{constraintstable_fermiondim7}{constraintstable_scalardim7},
the column denoted $\sigma_{\rm SI}$ carries a \yes~for models which
have zero hypercharge and are thus safe from direct detection, and
\no~otherwise.  For the latter, models of hypercharged DM can evade these direct detection if we allow for additional structures which we now discuss.

In particular, limits on spin-independent
scattering via the $Z$ boson are null if there is an even tiny mass splitting $\delta$ between
the components of the Dirac fermion or complex scalar
\cite{Hall:1997ah,TuckerSmith:2001hy}.  In this case, scattering
through the $Z$ boson is inelastic, requiring additional energy to excite the
incoming DM particle into a neighboring mass eigenvalue.  In particular,
scattering is kinematically forbidden provided the mass splitting exceeds
\begin{eqnarray} \delta \geq
\frac{\beta^2 \mu }{2}\, , \label{eq:splitting}
\end{eqnarray} 
where $\beta \sim 200$ km/sec is the DM velocity and $\mu$ is the
reduced mass of the DM-nucleus system.  For the typical atomic weights
of targets such in experiments like CDMS~\cite{Agnese:2013rvf},
XENON100~\cite{Aprile:2012nq}, and LUX~\cite{Akerib:2013tjd}, for $M
\sim 100$ GeV this translates into a bound $\delta \gtrsim 10$ keV.
For very small $\mdm$, the reduced mass will decrease and so too will the
lower bound on $\delta$, but these regions of light DM are excluded by
LEP for $Y \neq 0$~\cite{Heister:2002mn,Abdallah:2003xe}.

\begin{table}[t]
\begin{center}
Fermionic DM $(N=7)$
\end{center}
\begin{tabular}{|c|c|c|c|c|c|c|c|c|c|}
\hline
$SU(2)_L$ & $U(1)_Y$ & $U(1)_B$ & $U(1)_L$ & $\sslash{B},\sslash{L} $& $\sigma_{\rm SI}$  & ${\cal O}_{\rm DM}$ \\
\hline
 3 & 1 & $-1$ & 0 & \no & \no  & $H^\dagger Q \Dcbar^2 \chiDM$ \\
 3 & 1 & 0 & $-3$ & \no & \no  & $H L^3\chiDM$ \\
 4 & $-3/2$ & 0 & $-1$ &  \yess & \no  & $H^4 L\chiDM$ \\
 4 & 3/2 & 1 & 0 &  \yess & \no  & $H^\dagger \bar{Q}^2 \Uc \chiDM$ \\
 4 & $-1/2$ & 1 & 0 &  \yess & \no  & $H \bar{Q} ^2 \Dc \chiDM$ \\
 5 & $-1$ & $-1$ & 0 &  \yess  & \no  & $H Q^3\chiDM$ \\
 5 & 0 & $-1$ & 0 &  \yess & \yes  & $H^\dagger Q^3\chiDM$ \\
 5 & 1 & 0 & $-3$ & \yess & \no  & $H L^3\chiDM$ \\
 5 & 2 & 0 & $-3$ & \yess & \no  & $H^\dagger L^3\chiDM $ \\
 6 & $-3/2$ & 0 & $-1$ & \yes & \no  & $H^4 L\chiDM$ \\
 6 & $-1/2$ & 0 & $-1$ & \yes & \no  & $H^\dagger H^3 L\chiDM$ \\
 6 & 1/2 & 0 & $-1$ & \yes & \no  & $H^{\dagger 2} H^2 L\chiDM$ \\
 6 & 3/2 & 0 & $-1$ & \yes & \no  & $H^{\dagger 3} H L\chiDM$ \\
 6 & 5/2 & 0 & $-1$ & \yes & \no  & $H^{\dagger 4} L\chiDM$ \\

\hline
\end{tabular}
\caption{\label{table:constraintstable_fermiondim7} Classification of
  fermionic DM which is stable up to dimension seven decays.  Notation
  from \tableref{constraintstable_fermiondim6}, but with a \yess~to indicate models which are cosmologically stable at dimension five or less but may still decay at dimension six when $B$ and $L$ are violated consistent with proton stability.
  }
\end{table}

However, inducing the requisite mass splitting $\delta$ requires new
operators that explicitly break the DM particle number associated with
Dirac fermions or complex scalars.  For DM candidates that carry $B$
or $L$, this implies explicit breaking of $B$ or $L$.  However, the
spurion responsible for explicit breaking, $(\Delta B,\Delta L)$,
enters with twice the $B$ and $L$ charge of the DM. Consequently,
even with these splitting operators, there is still an unbroken
$Z_2$ subgroup of $B$ or $L$ that maintains DM stability.

For example, consider a fermionic DM particle that is a doublet of
hypercharge $Y=1/2$.  The leading operator that can split its
components is $H^{\dagger 2}\chiDM^2 /\Lambda$.  A mass splitting
sufficient to evade direct detection requires
\begin{eqnarray}
\Lambda \lesssim \frac{v^2}{\delta} \sim 10^{9}~\gev\,,
\label{eq:splittingLambda}
\end{eqnarray}
which is a low cutoff in the context of DM stability.  For even larger
hypercharges, the level splitting operator involves more Higgs fields,
and the requisite cutoff is even lower, dropping to $\sim$ 30 TeV for
$Y=1$.

According to the philosophy of effective theory, the bound in
\Eq{eq:splittingLambda} defines a cutoff scale at which all higher
dimension operators allowed by symmetry should be present---including
those which can mediate DM decay.  For dimension five, six, and seven
operators, such a low cutoff is inconsistent with cosmological limits
on weak scale DM.  This is depicted in \figref{LambdaLimits}, where
the blue shaded region shows that below above $\Lambda \gtrsim
10^{9}~\gev$, the cutoff is too high to induce a sufficient mass
splitting to evade direct detection, so $Y=0$ is forbidden.
Conversely, effectively stable DM with non-zero hypercharge can only
evade direct detection if there is a low cutoff, in which case
cosmological stability requires the DM decay operator be dimension
eight or higher.

\begin{table}[t]
\begin{center}
Scalar DM $(N=7)$
\end{center}

\begin{tabular}{|c|c|c|c|c|c|c|c|}
\hline
$SU(2)_L$ & $U(1)_Y$ & $U(1)_B$ & $U(1)_L$ & $\sslash{B},\sslash{L} $& $\sigma_{\rm SI}$  & ${\cal O}_{\rm DM}$ \\
 \hline
 1 & 0 & $-1$ & $-1$ &  \yess & \yes  & $Q^3 L  \chiDM$ \\
 2 & $-1/2$ & 1 & $-1$ & \no & \no  & $\Dc^3 L \chiDM$ \\
 2 & $-1/2$ & 0 & $-2$ & \no & \no  & $H^3 L^2 \chiDM$ \\
 2 & 1/2 & 1 & $-1$ & \no & \no  & $ \Uc \Dc^2 L \chiDM$ \\
 
 3 & $-1$ & $-1$ & $-1$ & \no & \no  & $ Q \bar\Uc^2 L \chiDM$ \\
 3 & $-1$ & 1 & $1$ & \no & \no  & $ \bar Q \Dc^2 \bar L \chiDM$ \\
 3 & 0 & $-1$ & $-1$ & \yess & \yes  & $ Q^3 L \chiDM$ \\

 4 & $-3/2$ & $-1$ & 1 & \no & \no  & $ Q^3 \Ec \chiDM$ \\
  4 & $-1/2$ & $-1$ & 1 & \no & \no  & $ Q^2 \bar \Uc \bar L \chiDM$ \\
 4 & $-1/2$ & 0 & $-2$ & \no & \no  & $H^3 L^2 \chiDM$ \\

 5 & 0 & $-1$ & $-1$ & \yess & \yes  & $ Q^3 L \chiDM$ \\
 5 & 2 & 0 & $-4$ & \no & \no  & $L^4 \chiDM$ \\

 6 & $-1/2$ & 0 & $-2$ &  \yess & \no  & $H^3 L^2 \chiDM $ \\
 6 & 1/2 & 0 & $-2$ &  \yess & \no  & $H^\dagger H^2 L^2 \chiDM $ \\
 6 & 3/2 & 0 & $-2$ & \yess & \no  & $H^{\dagger 2} H L^2 \chiDM $ \\
 6 & 5/2 & 0 & $-2$ &  \yess & \no  & $H^{\dagger 3} L^2 \chiDM $ \\

 7 & 0 & 0 & 0 & \yes & \yes  & $H^{\dagger 3} H^3 \chiDM$ \\
 7 & 1 & 0 & 0 & \yes & \no  & $H^{\dagger 4} H^2 \chiDM$ \\
 7 & 2 & 0 & 0 & \yes & \no  & $H^{\dagger 5} H \chiDM$ \\
 7 & 3 & 0 & 0 & \yes & \no  & $H^{\dagger 6} \chiDM$ \\

\hline
\end{tabular}
\caption{\label{table:constraintstable_scalardim7} Classification of
  scalar DM which is stable up to dimension seven decays.  Notation from
  \tableref{constraintstable_fermiondim6}.}

\end{table}

In general, if the mass splitting requires a higher dimension operator
then direct detection is inconsistent with the criterion of
cosmologically stable DM.  On the other hand, there is no issue if the
splitting operator is renormalizable.  For example, this is possible
for complex scalar DM with $Y= 1/2$, which permits the
renormalizable operator $\lambda H^{\dagger 2} \chiDM^2 $.  For
$\mdm=100~\gev$, a splitting of $\delta = 10~\kev$ is achieved for
$\lambda \gtrsim 10^{-9}$, which is easily satisfied for order one couplings.
If $\chiDM$ carries $B$ or $L$, this mass splitting operator
explicitly breaks $B$ or $L$ down to a discrete baryon or lepton
parity, however still maintaining DM stability.  Alternatively, this
interaction is symmetry preserving if $\chiDM$ does not carry $B$ or
$L$, as is the case for $k$-plets of $SU(2)_L$ with large values of
$k$ which couple only to Higgs bosons in the decay operator.

Even for models which evade bounds on $Z$-mediated scattering, direct
detection may still impose constraints on Higgs boson exchange.  For
example, scalar DM-nucleon scattering can occur via the
Higgs portal interaction, $H^\dagger H X^\dagger
X$~\cite{Schabinger:2005ei}.  For fermionic DM, the analogous
interactions are higher dimension.  Non-singlet fermionic and scalar
candidates can also scatter with nucleons at loop level via multiple
gauge boson exchange, which may be observable in the next generation
of direct detection experiments~\cite{Cline:2013gha}.

\subsection{Indirect Detection}

Since these DM candidates eventually decay on cosmological time scales, they are naturally
probed by cosmic ray telescopes.  Conveniently, the authors of
\cite{Zhao:2014nsa} studied indirect
detection constraints on DM decay via high dimension operators of the
very type considered in this paper, albeit with underlying motivation
of asymmetric DM. In particular, they considered bounds from FERMI, PAMELA,
AMS-02, and HESS on high energy gamma rays and charge particle cosmic
rays from electrons, protons, and anti-protons, obtaining a limit on
the DM lifetime which we have taken as a loose input for \Eq{eq:ID}.
We refer the reader to \cite{Zhao:2014nsa} for precise numerical
bounds, but we summarize the salient takeaways below.

In general, DM carrying $B$ will decay to quarks, yielding
anti-protons, while DM carrying $L$ will decay to leptons, yielding
positrons and neutrinos.  All of the DM decays considered here will
produce high energy gamma rays from charged particle bremsstrahlung,
hadronic decays, and inverse Compton scattering of CMB photons and
starlight.  However, due to the large $B$ and $L$ charges of most our
DM candidates, gamma ray lines are not typically expected among the
theories considered here.  An exception occurs for DM particles which
decay through operators involving only the Higgs boson, in which case
mixing together with a loop of SM particles will induce two-body
decays of DM to photons.  Another possibility occurs for DM with unit
$L$ number, which can decay to photon plus neutrino.

\subsection{Proton Decay}

The non-observation of proton decay offers a strong motivation for at
least approximate $B$ and $L$ conservation.  Current limits on $p^+
\rightarrow e^+ \pi^0$ and related decay modes from the
Super-Kamiokande experiment require a lifetime of at least $\sim
10^{33}$
years~\cite{Abe:2014mwa,Takhistov:2014pfw,Abe:2013lua,Regis:2012sn},
which already places significant constraints on the simplest GUTs. From an effective theory
viewpoint, proton decay is mediated by dimension six operators of
the form
\begin{eqnarray}
\frac{Q^3L}{\Lambda^2},
\label{eq:QQQL}
\end{eqnarray}
which breaks $B$ and $L$ but preserves $B-L$.  Current limits on
proton decay imply a lower bound on the cutoff of approximately
$\Lambda \gtrsim 10^{15}$~GeV~\cite{Nath:2006ut}, so $B$ and $L$ are very well-preserved symmetries.

As noted earlier, we can parameterize $B$ and $L$ violation with a dimensionless
spurion characterizing hard breaking by some number of units of $\Delta
B$ and $\Delta L$.  However, $(\Delta B,\Delta L)= (1,1)$ is of
particular note because it permits proton decay via the operator in
\Eq{eq:QQQL}.  As a result, if we wish to forbid this then we should avoid breaking by units of $(\Delta
B,\Delta L)= (1,1)$.  Remarkably, even with $(\Delta B,\Delta L)\neq
(1,1)$ there are still viable models of effectively stable DM.  In
these models there are low dimension DM decay operators allowed by the
SM gauge symmetry, but these operators require explicit breaking by
$(\Delta B,\Delta L)= (1,1)$ which would also decay the proton. 

For example, this happens in the model described by the first row of \tableref{constraintstable_fermiondim6}.  Since $X$ is a gauge singlet, the gauge invariant operator $HLX$, if present, would induce catastrophically prompt DM decay.  However, the existence of both $UD^2 X$ together with $HLX$ would require symmetry breaking by $(\Delta B, \Delta L)=(1,1)$ which would in turn induce dimension six proton decay.  Conversely, if $B$ and $L$ are
explicitly broken while forbidding proton decay, an accidental symmetry remains which effectively stabilizes DM.

Note that proton decay is also mediated by operators beyond dimension six requiring $B$ and $L$ violation by one unit and an odd number of units, respectively.  For example, $H^\dagger \Dc^3 L$ induces proton decay with $(\Delta B,\Delta L) = (-1,1)$ breaking.
Forbidding this larger class of spurions would allow for even more viable candidates for cosmologically stable DM, but we do not consider this possibility here.

In
\tablesref{constraintstable_fermiondim6}{constraintstable_scalardim6}
and
\tablesref{constraintstable_fermiondim7}{constraintstable_scalardim7},
the column labelled $\sslash B, \sslash L$ carries a \yes~if DM is
still effectively stable---that is, cannot decay at dimension five or
less---even after including any hard breaking spurion with $(\Delta B, \Delta
L) \neq(1,1)$, and \no~otherwise.  
The \yess~entries in \tablesref{constraintstable_fermiondim7}{constraintstable_scalardim7} indicate models with DM that, while stable at dimension five or less, still decays at dimension six.  For these models, DM is still cosmologically stable if the cutoff is higher than was required for dimension seven decays.

\section{Summary and Outlook}

\label{sec:conclusion}

DM phenomenology often hinges on the assumption of a stabilizing
symmetry.  
Naturally, this leads one to wonder about the
underlying reason for cosmological stability.
In this paper we present an alternative hypothesis whereby DM is
long-lived as an accident of the SM symmetry group.  We have
classified all models in which DM decay is cosmologically slow and
induced by dimension six or seven operators.  In such cases a
sub-Planckian cutoff is sufficient to prevent cosmologically prompt DM
decays which are excluded by indirect detection. All candidates for effectively stable DM either carry $B$ or $L$, or reside in a high representation
of $SU(2)_L$.  We have identified those models which are consistent
with stringent bounds on spin-independent DM-nucleon scattering.
Finally, we have accounted for the possibility explicit breaking of
$B$ and $L$ by arbitrary units.  As long as the symmetry breaking
spurion still forbids proton decay, models of effectively stable DM persist.

Our analysis leaves a number of avenues for future work.  First, there is
the question of DM stabilization by SM quark and lepton flavor symmetries, which are
well-preserved for the light generations.  Second, there is the question of building these models in a supersymmetric context.  Lastly, it would be
interesting to see if any of the models presented here 
arise explicitly in GUT constructions.

\section*{Acknowledgments}

CC is supported by a DOE Early Career Award
DE-SC0010255 and a Sloan Research Fellowship.  DS is supported in part by U.S.~Department of Energy
grant DE--FG02--92ER40701 and by the Gordon and Betty Moore Foundation
through Grant No.~776 to the Caltech Moore Center for Theoretical
Cosmology and Physics.

\bibliography{bibeffdecay_01}{}

\end{document}